\documentclass[rnote]{aa}
\bibpunct{(}{)}{;}{a}{}{,} 
\usepackage{graphicx}
\begin{document}
\title{A curious relation between the flat cosmological model and the elliptic integral of the first kind}
\titlerunning{A curious relation between the ...}  

\author{A. M\'esz\'aros \inst{1}
\and
J. \v{R}\'{\i}pa \inst{2}
}

\offprints{A. M\'esz\'aros}

\institute{Charles University in Prague, Faculty of Mathematics and Physics, Astronomical Institute,\\
V Hole\v{s}ovi\v{c}k\'ach 2, CZ 180 00 Prague 8, Czech Republic\\
\email{meszaros@cesnet.cz}\\
\and
Institute of Basic Science, Natural Sciences Campus, Sungkyunkwan University\\
Engineering Building 2, 2066 Seobu-ro, Jangan-gu, Suwon, Gyeonggi-do, 440-746, Korea\\
\email{ripa.jakub@gmail.com}
}

\date{Received June 14, 2013; accepted June 16, 2013}

\abstract {The dependence of the luminosity distance on the redshift
has a key importance in the cosmology. This dependence
can well be given by standard functions for the zero cosmological constant.}
{The purpose of this article is to present such a relation also for the non-zero cosmological constant, 
if the universe is spatially flat.}
{A definite integral is used.}
{The integration ends in the elliptic integral of the first kind.}
{The result shows that no numerical integration is needed for the non-zero cosmological constant, 
if the universe is spatially flat.}

\maketitle

\section{Introduction}

The dependence of the luminosity distance $d_L(z)$ on the redshift $z$ 
is a key formula in cosmology \citep{car92}.
It is given by three independent cosmological parameters: by two omega-parameters 
$\Omega_M$, $\Omega_{\Lambda}$ and by the Hubble constant $H_{0}$. 
Its relation to the so called "proper-motion distance" is given by
 $d_{PM}(z) (1+z) = d_L(z)$ \citep{wei72}.  

One has \citep{car92}:
$$ d_{PM}(z) = \frac{c}{H_0 \sqrt{|\Omega_k|}} \times \;\;\;\;\;\;\;\;$$
\begin{equation}
  {\rm sinn} 
 \left\{\sqrt{|\Omega_k|} \int_{0}^{z}
                   \frac{dz'}{\sqrt{(1+z')^2 (1+\Omega_M z') - z'(2+z')\Omega_{\Lambda}}}\right\}.
\end{equation}
In this equation $c$ is the speed of light in vacuum, 
and it holds $\Omega_k + \Omega_M + \Omega_{\Lambda} = 1$. The notation "sinn" means the standard function
$\sinh$ for $\Omega_k > 0$, and $\sin$ for $\Omega_k < 0$, respectively. If $\Omega_k = 0$, then one simply has
an integration: 
\begin{equation}
\frac{H_{0} d_{PM}(z)}{c} = 
                   \int_{0}^{z} \frac{dz'}{\sqrt{(1+z')^2 (1+\Omega_M z') - z'(2+z')\Omega_{\Lambda}}}.
\end{equation}

In addition, from the physical point of view, in both equations it must be $\Omega_M > 0$. 
(The case  $\Omega_M = 0$ can serve as a limit, but $\Omega_M < 0$ is fully unphysical.)
On the other hand, $\Omega_{\Lambda}$ can have both signs, but the observations of the last 
two decades strongly disfavour negative values (see, e.g., \citet{per12} 
and the references therein).

In the special case of $\Omega_{\Lambda} = 0$ the integral in Eq.(1) can be given by 
the so-called Mattig-formula \citep{mat58} for any $\Omega_M > 0$. 
The formula can then well be used in
cosmological applications (see, e.g., \citet{mesz02}). 
For $\Omega_{\Lambda} \neq 0$ the integral in Eqs.(1,2) 
is usually solved 
numerically\footnote{See, cf., http://www.astro.ucla.edu/$\sim$wright/cosmolog.htm}. 

In this note we show that the integral on the right-hand-side of Eq.(2) 
can be solved analytically also for $\Omega_{\Lambda} \neq 0$.

\section{The integration}

We rewrite the right-hand side of Eq.(2) into the form 
\begin{equation}
\frac{H_{0} d_{PM}(z_0)}{c} = I(\Omega_M, z_0) =  \int_{0}^{z_0} \frac{dz}{\sqrt{(1+z)^3 \Omega_M  + \Omega_{\Lambda}}},
\end{equation}
where $z \geq 0$, $\Omega_M > 0$ and  $\Omega_M + \Omega_{\Lambda} = 1$. 
Introducing the substitution $x = (1+z)^{-1}$, we obtain
\begin{equation}
I(\Omega_M, x_0) =  \int_{x_0}^{1} \frac{dx}{\sqrt{x} \sqrt{\Omega_M  + \Omega_{\Lambda}x^3}},
\end{equation}
where $0 < x_0 = (1+ z_0)^{-1} \leq 1$.

In this section we consider only the case $\Omega_M  < 1$, i.e. $\Omega_{\Lambda} = 1 - \Omega_M > 0$.
Introducing a further substitution $y = (\Omega_{\Lambda}/\Omega_M)^{1/3} x$,
we obtain
\begin{equation}
I(\Omega_M, x_0) =  \frac{1}{\Omega_M^{1/3} \Omega_{\Lambda}^{1/6}}
\int_{y_1}^{y_2} \frac{dy}{\sqrt{y} \sqrt{1+ y^3}},
\end{equation}
where $y_1 = (\Omega_{\Lambda}/\Omega_M)^{1/3} x_0$ and $y_2 = (\Omega_{\Lambda}/\Omega_M)^{1/3}$.
The two limits are non-negative with $y_1 \leq y_2$. 

We try to find the primitive function for 
\begin{equation}
\int \frac{dy}{\sqrt{y} \sqrt{1+ y^3}},
\end{equation}
where $y \geq 0$.
The following substitution helps:
\begin{equation}
\alpha =  \arccos \left(\frac{1+ (1-\sqrt{3}) y}{1+ (1+\sqrt{3}) y} \right).
\end{equation}
There is a one-to-one correspondence between $\alpha$ and $y$. For $y=0$ one has $\alpha = 0$, and
for $y \rightarrow \infty$ one has $\alpha_{\infty} \rightarrow \arccos((1 - \sqrt{3})/(1+\sqrt{3}))$, 
i.e. $\alpha_{\infty} = 105.54^{\circ}$.                          
If $y$ increases in the interval $[0,\infty)$, $\alpha$ is 
also increasing in the interval $[0, \alpha_{\infty})$. Hence, this 
substitution is well-defined. Conversely, one obtains
\begin{equation}
y = \frac{1 - \cos \alpha}{(\sqrt{3} -1) + (\sqrt{3} + 1) \cos \alpha}
\end{equation}
and
\begin{equation}
d y = \frac{2\sqrt{3} \sin \alpha \; d\alpha}{[\sqrt{3}(1+\cos \alpha) - (1-\cos \alpha)]^2}.
\end{equation}
Using these two formulas, we curiously obtain
\begin{equation}
\frac{dy}{\sqrt{y} \sqrt{1+ y^3}} =  \frac{d \alpha }{3^{1/4} \sqrt{1 - \frac{2 +\sqrt{3}}{4} \sin^2 \alpha}}. 
\end{equation}

The right-hand side of Eq.(10) is the function 
in the elliptic integral of the first kind
\citep{gra07} with
\begin{equation}
m = \frac{2 +\sqrt{3}}{4}, 
\end{equation}
where $0 < m <1$, as it should be in an elliptic integral.

In order to calculate the definite integral 
\begin{equation}
\int_{y_1}^{y_2} \frac{dy}{\sqrt{y} \sqrt{1+ y^3}}
\end{equation}
in Eq.(5) for non-negative $y_1 \leq y_2$ one can write
\begin{equation}
\int_{y_1}^{y_2} \frac{dy}{\sqrt{y} \sqrt{1+ y^3}} = 
\int_{0}^{y_2} \frac{dy}{\sqrt{y} \sqrt{1+ y^3}} - 
\int_{0}^{y_1} \frac{dy}{\sqrt{y} \sqrt{1+ y^3}}. 
\end{equation}
After this one should use the formula with $\alpha$ from Eq.(10) and determine the integration limits
in variable $\alpha$. The substitution from Eq.(7) gives for $y=0$ the value $\alpha = 0$. This means
that  - using $\alpha$ - the lower limits in both definite integrals are zeros. The upper limits from $y_1$ and $y_2$,
respectively, are also calculable analytically and 
unambiguously from Eq.(7) via the $\arccos$ function. One obtains $\alpha_1$ and 
$\alpha_2$, respectively, as upper limits in the integrals. It must be $\alpha_2 > \alpha_1$. 
One should only precise that, of course, if it were $\alpha_2$ 
from the interval $(\pi/2, \alpha_{\infty})$, then the first elliptic integral itself should be given by a sum of 
two integrals: In one integral the limits should be $0$ and $\pi/2$, and in the second one the limits should
be $(\pi - \alpha_2)$ and $\pi/2$. Both integrals must give positive values.
If it were also $\alpha_1 > \pi/2$, then one should proceed similarly in the second integral, too.  
In any case, the integral $I(\Omega_M, x_0)$ is well obtainable from standard elliptic integrals of the first kind.

\section{Remarks}

Integral in Eq.(5) is presented by \citet{gra07} (formula 3.166.22). 
It should also be noted that in \citet{car92} 
it is said that the integral of Eq.(2) can also be solved analytically.
In \citet{paal92} similar efforts are done. But, on the other hand, 
we did not find in the literature any note about this non-numerical integration of Eq.(2)
using the elliptic integrals. Therefore, we mean that the substitution given by Eqs.(7,8) is new and original.

For the sake of completeness it should still be added that integral of Eq.(4) can be solved also for 
$\Omega_{\Lambda} <0$ and $\Omega_M > 1$. One obtains (up to a constant) the formula $dy/\sqrt{y(1- y^3)}$, which 
is also integrable (see the formula 3.166.23 of \citet{gra07}).
  
\section{Conclusion}

We have proven that the integral on the right-hand-side of Eq.(2) 
can be solved also analytically using the elliptic integral of the first kind.

\begin{acknowledgements}
We wish to thank M. K\v{r}\'{\i}\v{z}ek for the useful discussions and comments on the manuscript.
This study was supported by the OTKA Grant K77795, by the Grant Agency of the Czech Republic Grant
P209/10/0734, by the Research Program MSM0021620860 of the Ministry of Education of the Czech Republic,
and by Creative Research Initiatives (RCMST) of MEST/NRF.
\end{acknowledgements}

\bibliographystyle{aa} 
\bibliography{references-elliptic}

\end{document}